\documentclass[aps,prd,nofootinbib,twocolumn,floatfix,superscriptaddress]{revtex4-1}
\usepackage{setspace}
\usepackage{tabularx, booktabs}
\newcolumntype{Y}{>{\centering\arraybackslash}X}
\usepackage{graphicx}
\usepackage{dcolumn}
\usepackage{multirow}
\usepackage{subfigure}
\usepackage{times,mathptm}
\usepackage{float}
\usepackage{color}
\usepackage{amsmath,amsfonts}
\usepackage{mathptmx}
\usepackage{mathrsfs}
\usepackage{bbm}
\usepackage{bm}
\usepackage{xfrac}

\newcommand{\beq}{\begin{equation}}
\newcommand{\eeq}{\end{equation}} 
\newcommand{\bea}{\begin{eqnarray}}
\newcommand{\eea}{\end{eqnarray}}

\renewcommand{\P}{\overline{P}}
\renewcommand{\d}{\delta}

\renewcommand{\l}{\lambda}

\newcommand{\Dt}{{\cal D}}

\newcommand{\tK}{\widetilde{K}}
\newcommand{\G}{G_{\infty}}
\newcommand{\pbar}{\overline{\psi}}

\renewcommand{\b}{\beta}
\renewcommand{\a}{\alpha}

\newcommand{\tr}{\text{Tr}}

\newcommand{\vx}{{\vec{x}}}
\newcommand{\vy}{{\vec{y}}}

\newcommand{\vk}{{\vec{k}}}

\newcommand{\m}{\mu}

\renewcommand{\k}{\kappa}

\newcommand{\D}{\Delta}

\newcommand{\oh}{\frac{1}{2}}

\newcommand{\on}{\frac{1}{9}}
\newcommand{\dg}{\dagger}
\newcommand{\non}{\nonumber}
\renewcommand{\t}{\tau}
\newcommand{\rf}[1]{(\ref{#1})}
\newcommand{\ra}{\rightarrow}
\newcommand{\pa}{\partial}
\renewcommand{\vec}[1]{\bm #1}

\usepackage{ulem}

\begin{document}

\title{A finite-density transition line for QCD with 695 MeV dynamical fermions} 
 
\author{Jeff Greensite}
\affiliation{Physics and Astronomy Department, San Francisco State
University,   \\ San Francisco, CA~94132, USA}
\bigskip
\author{Roman H\"ollwieser}
\affiliation{Theoretical Particle Physics, Bergische Universit\"at Wuppertal, \\ Gau{\ss}str.\ 20, 42119 Wuppertal, Germany}
\date{\today}
\vspace{60pt}
\begin{abstract}

\singlespacing

    We apply the relative weights method to SU(3) gauge theory with staggered fermions of mass 695 MeV at a set of temperatures
in the range $151 \le T \le 267$ MeV, to obtain an effective Polyakov line action at each temperature.  We then apply a mean-field 
method to search for phase transitions in the effective theory at finite densities.  The result is a transition line in the plane of temperature and chemical potential, with an endpoint at high temperature, as expected, but also a second endpoint at a lower temperature.  We cannot rule out the possibilities that a transition line reappears at temperatures lower than the range investigated, or that the second endpoint is absent
for light quarks.
 
\end{abstract}

\maketitle

\singlespacing

\section{\label{intro}Introduction}

     The effective Polyakov line action (PLA) of a lattice gauge theory is the theory which results from integrating out all of the degrees
of freedom of the theory, subject to the condition that the Polyakov lines are held fixed, and it is hoped that this effective theory is
more tractable than the underlying lattice gauge theory (LGT) when confronting the sign problem at finite density.  The general idea was pioneered in \cite{Fromm:2011qi}, and the derivation of the PLA from the underlying LGT has been pursued by various methods, e.g.\ \cite{Gattringer:2011gq,Bergner:2015rza,Wozar:2007tz}.  The relative weights method \cite{Greensite:2014isa,*Greensite:2013bya} is a simple numerical technique for finding the derivative of the PLA in any direction in
the space of Polyakov line holonomies.\footnote{The term ``holonomy'' refers here to the group element corresponding to a closed Wilson line, prior to taking the trace.}  Given some ansatz for the PLA, depending on some set of parameters, we can use the
relative weights method to determine those parameters.  Then, given the PLA at some fixed temperature $T$, we can apply a mean
field method to search for phase transitions at finite chemical potential $\m$.  This is the strategy which we have outlined in some
detail in \cite{Hollwieser:2016hne}, where some preliminary results for finite densities were presented.  The relative weights method has strengths and weaknesses; on the positive side the approach is not tied to either a strong coupling or hopping parameter expansion, and the
non-holomorphic character of the fermion action is irrelevant.  The main
weakness is that the validity of the results depends on a good choice of ansatz for the PLA.  We have suggested, for exploratory work, an ansatz for the PLA inspired first by the success of the relative weights method applied to pure gauge theories 
\cite{Greensite:2014isa,*Greensite:2013bya}, and secondly by the form of the PLA obtained for heavy-dense quarks.

    In this article we follow up on the work in ref.\  \cite{Hollwieser:2016hne} to obtain a tentative transition line in the $\m-T$ plane for SU(3) gauge theory with dynamical staggered unrooted quarks of mass 695 MeV.  It is generally believed that this line has an endpoint at high temperatures, and this is what we find.  A second, unexpected finding is that there is also an endpoint at a lower temperature.\footnote{In fact some conjectured QCD phase diagrams do contain a second endpoint (see, e.g.\ \cite{Fukushima:2010bq,Sasaki:2009zh}), but this is based on the idea of quark hadron continuity at $N_f=3$ flavors \cite{Schafer:1998ef}, and it is not clear that a similar argument would apply in our case, with unrooted staggered fermions.} 
Whether a transition line reappears at still lower temperatures, outside the range we have investigated, or whether the second transition point disappears for lighter quarks, or whether this second transition point is instead indicative of some deficiency in our ansatz for the PLA, remains to be seen.  

    In the next section we briefly review the relative weights method and associated mean field technique at finite density, referring to our previous work \cite{Hollwieser:2016hne}  for some of the technical details.  Section 3 contains our results, followed by conclusions in section 4.  A result for a recently introduced observable $\xi/\xi_{2nd}$ \cite{Caselle:2017xrw}, which is sensitive to excitations above the lowest
lying excitation, is presented in an appendix.

\section{\label{review}Relative Weights}
    
    It is simplest to work in temporal gauge, where we can fix the timelike links to the identity everywhere except on one timeslice,
say at $t=0$, on the periodic lattice.  In this gauge the timelike links at $t=0$ are the Polyakov line holonomies, which are held fixed, and the PLA $S_P$ for an SU(N) gauge theory is defined by
\bea
\lefteqn{\exp\Bigl[S_P[U_{\vx},U^\dg_{\vx}]\Bigl] }& & \non \\
&=&  \int  DU_0(\vx,0) DU_k  D\psi D\pbar \left\{\prod_{\vx} \d[U_{\vx}-U_0(\vx,0)]  \right\} e^{S_L} \ , \non \\
\label{S_P}
\eea
where $S_L$ is the lattice action for an SU(N) gauge theory with dynamical fermions, and we also define the Polyakov line 
$P_\vx = {1\over N} \tr U_\vx$.  In this article we use the standard SU(3) Wilson action for the gauge field, and unrooted staggered fermions as the dynamical matter fields.  Given the PLA at $\m=0$, the PLA at finite $\m$ is obtained by the simple replacement
\bea
     S_P^\m[U_\vx,U^\dg_\vx] =  S_P^{\m=0}[e^{N_t \m} U_\vx,e^{-N_t \m}U^\dg_\vx]   \ ,
\label{convert}
\eea
where $N_t$ is the lattice extension in the time direction, with $\m$ in lattice units.
 
     Let us consider some path through the space of Polyakov line holonomies $U_\vx(\l)$ parametrized by $\l$.  The relative weights
method allows us to compute the derivative $dS_P/d\lambda$ anywhere along the path.  Let $U'_\vx$ and $U''_\vx$ represent
the Polyakov line holonomy field at parameter $\l_0 \pm \oh \D \l$ respectively.  Defining $S'_L, S''_L$ as the lattice actions
with the Polyakov line holonomies fixed to $U'_\vx, U''_\vx$ respectively, and $ \D S_P = S_P[U'_\vx] - S_P[U''_\vx]$, then in temporal gauge we have by definition
\bea
e^{\D S_P} &=&  {\int  DU_k D\psi D\pbar ~  e^{S'_L} \over \int  DU_k  D\psi D\pbar ~  e^{S''_L} }
\non \\ 
&=& {\int  DU_k  D\psi D\pbar ~  \exp[S'_L-S''_L] e^{S''_L} \over \int  DU_k  D\psi D\pbar ~  e^{S''_L} }
\non \\
&=& \Bigl\langle  \exp[S'_L-S''_L] \Bigr\rangle'' \ ,
\label{rw}
\eea
where $\langle ... \rangle''$ indicates that the expectation value is to be taken in the probability measure
\bea
{e^{S''_L} \over  \int  DU_k  D\psi D\pbar ~  e^{S''_L} } \ .
\eea
The expectation value in the last line of \rf{rw} can be calculated by standard lattice Monte Carlo, only holding fixed timelike
links at $t=0$.    From this calculation we find the derivative $dS_P/d\lambda \approx \D S_P / \D \lambda$.

   The SU(2) and SU(3) gauge groups are special in the sense that the trace $P_\vx$ determines the holonomy $U_\vx$.  So
we expand
\beq
             P_\vx = \sum_\vk  a_\vk e^{i \vk \cdot \vx} \ ,
\eeq
and compute the derivatives $\pa S_P/ \pa a_\vk$ with respect to a set of specific Fourier components $a_\vk$, keeping the remaining components fixed to values taken from a thermalized configuration.  Our ansatz for the PLA  is
\bea
e^{S_P} &=& \left(\prod_\vx \det[1+he^{\mu/T}\tr U_\vx]^p\det[1+he^{-\mu/T}\tr U^\dg_\vx]^p \right) \non \\
& & \times \exp\left[  \sum_{\vx \vy} P_\vx P^\dg_\vy K(\vx-\vy)  \right] \ ,
\eea
where $p=1$ for unrooted staggered fermions, and $p=2N_f$ for Wilson fermions, where $N_f$ is the number of flavors. The determinant factors
\bea
\lefteqn{\det[1+he^{\mu/T}\tr U_\vx]} & &  \non \\
&=&1+he^{\mu/T}\tr U_\vx+h^2e^{2\mu/T}\tr U^\dg_\vx+h^3e^{3\mu/T}, \label{det1} \\
\lefteqn{\det[1+he^{-\mu/T}\tr U^\dg_\vx]} & & \non \\
 &=& 1+he^{-\mu/T}\tr U^\dg_\vx+h^2e^{-2\mu/T}\tr U_\vx+h^3e^{-3\mu/T} 
\label{det2}
\eea
are motivated by the PLA for heavy dense quarks \cite{Fromm:2011qi,Bender:1992gn,*Blum:1995cb,*Engels:1999tz,*DePietri:2007ak} in which $h = (2\kappa)^{N_t}$, with $\kappa$ the hopping parameter for Wilson fermions, or $\kappa=1/2m$ for staggered fermions.  In our ansatz, $h$ becomes a fit parameter.  The part of the action involving
the kernel $K(\vx-\vy)$ is motivated by previous successful treatments \cite{Greensite:2014isa,*Greensite:2013bya} of pure gauge theory and gauge-Higgs theory.  All in all, using
\beq
\sum_{\vx,\vy} P_\vx K(x-y) P^\dg_\vy = \sum_\vk a_k a^*_k \tK(\vk) \ ,
\eeq
where
\beq
             K(\vx-\vy) = {1\over L^3} \sum_\vk \tK(\vk) e^{-\vk \cdot (\vx-\vy)} 
\eeq
on an $L^3$ three-space volume, and choosing $p=1$, we have 
\begin{widetext}
\bea
S_P[U_\vx] &=& \sum_\vk a_k a^*_k \tK(\vk)
+ \sum_\vx \bigg\{ \log(1+he^{\mu/T}\tr[U_\vx]+h^2e^{2\mu/T}\tr[U_\vx^\dagger]+h^3e^{3\mu/T}) \non \\
& & \qquad +\log(1+he^{-\mu/T}\tr[U_\vx]+h^2e^{-2\mu/T}\tr[U_\vx^\dagger]+h^3e^{-3\mu/T}) \bigg\} \ .
\label{eq:SP}
\eea
\end{widetext}

   The Fourier transform $\tK(\vk)$ of $K(\vx-\vy)$ is determined numerically at $\m=0$.  For $\vk=0$,
\bea
\lefteqn{{1\over L^3} \left( {\pa S_P \over \pa a^R_0} \right)_{a_0=\a}} & & \non \\
&=& 2 \tK(0) \a 
 + \bigg\{(3 h   + 3h^2  ) {1\over L^3}\sum_\vx Q_\vx^{-1} + \mbox{c.c} \bigg\} \ ,
\label{K0a}
\eea
where
\beq
          Q_\vx = 1 + 3 h P_\vx+  3h^2   P_\vx^\dg + h^3  \ .
\label{Q}
\eeq
If $h \ll 1$, then dropping terms of $O(h^2)$ and higher the derivative simplifies to
\bea
{1\over L^3} \left( {\pa S_P \over \pa a^R_0} \right)_{a_0=\a} = 2 \tK(0) \a + 6 h  \ ,
\label{deriv1}
\eea
where the ``R'' superscript refers to the real part.
For $\vk \ne 0$, again dropping terms of order $h^2$ and higher,
\bea
 {1\over L^3} \left( {\pa S_P \over \pa a^R_\vk} \right)_{a_\vk=\a} = 2 \tK(\vk) \a \ .
\label{deriv2}
\eea

The derivatives on left-hand sides of \rf{deriv1} and \rf{deriv2} are determined by the method of relative weights at a variety of $\a$.  By plotting those results vs.\ $\a$ and fitting the data to a straight line, $K(\vk)$ is determined from the slope.   The $h$ parameter can in principle be determined by the $y$-intercept of the $K(0)$ vs.\ $\a$ data.   A better method, for reasons to be discussed below, is to choose $h$ by requiring that $\langle P \rangle$ computed in the PLA at $\m=0$ agrees with $\langle P \rangle$ computed in the underlying LGT, 
\begin{figure}[htb]
 \includegraphics[scale=0.6]{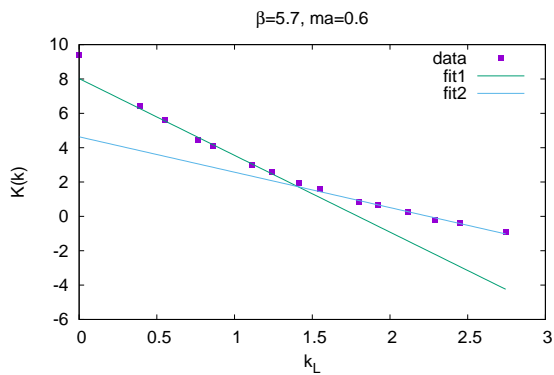}
\caption{Fits to the relative weights results for $\tK(\vk)$ vs $k_L$, obtained at $\b=5.7,~ma=0.6,~N_t=6$ by the double line fit \rf{Kfit1}.}  
\label{Kfit}
\end{figure}

\begin{figure}[h!]
\subfigure[]  
{\includegraphics[scale=0.6]{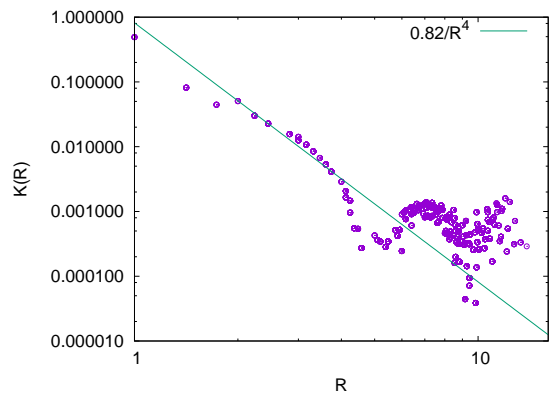}
\label{KR}
}
\subfigure[]  
{\includegraphics[scale=0.6]{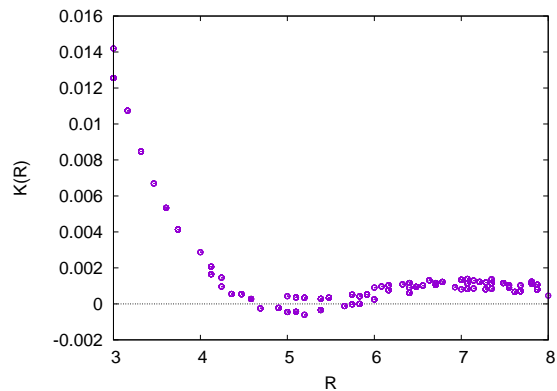}
\label{KRsmall}
}
 \caption{Finite Fourier transform of the double-line fit for $\tK(k)$ to position space $K(R)$. (a) logarithmic plot; (b) linear plot in the
 region $3<R<8$.}
\label{KRall}
\end{figure}

\section{Deriving the Polyakov loop action}

\begin{figure}[htb]
\subfigure[~$\b=5.70,~ma=0.6$]  
{   
 \includegraphics[scale=0.6]{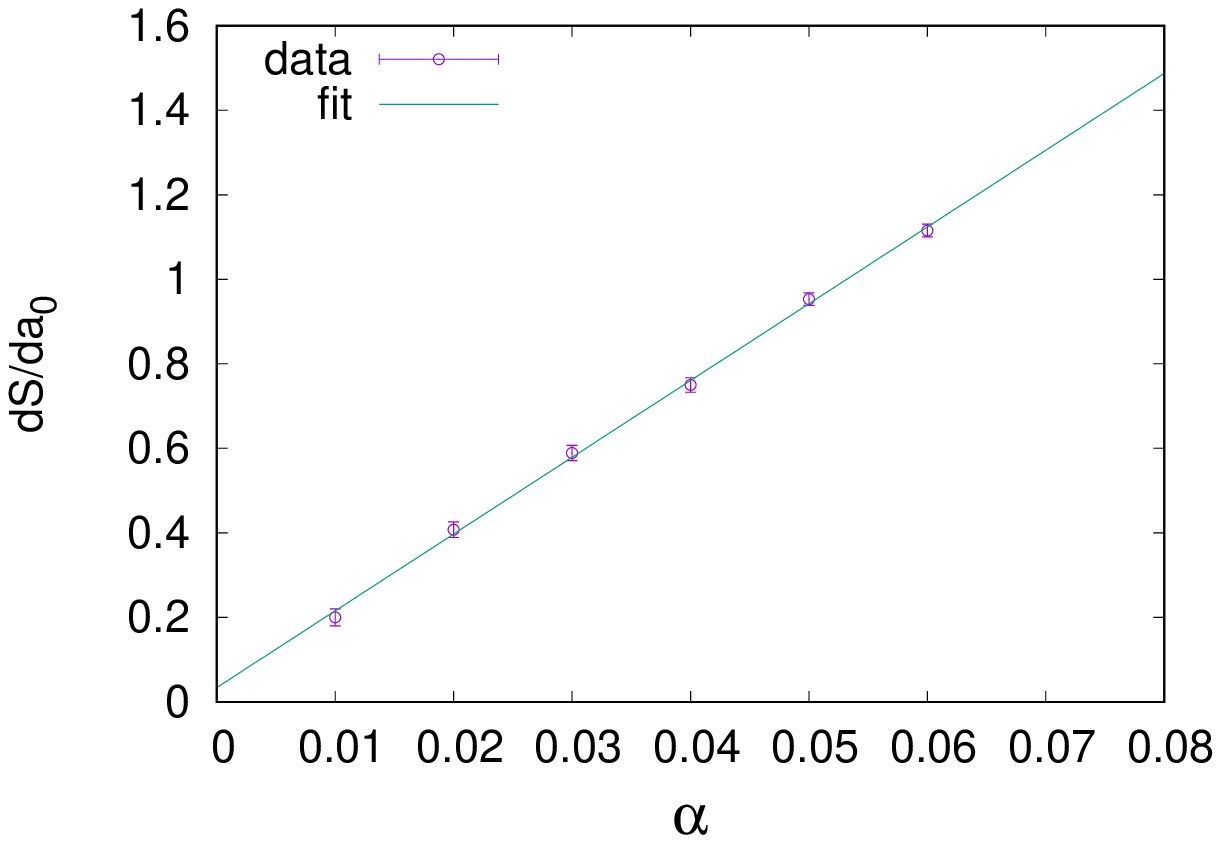}
 \label{dSa}
}
\subfigure[~$\b=5.85,~ma=0.435$]  
{   
 \includegraphics[scale=0.6]{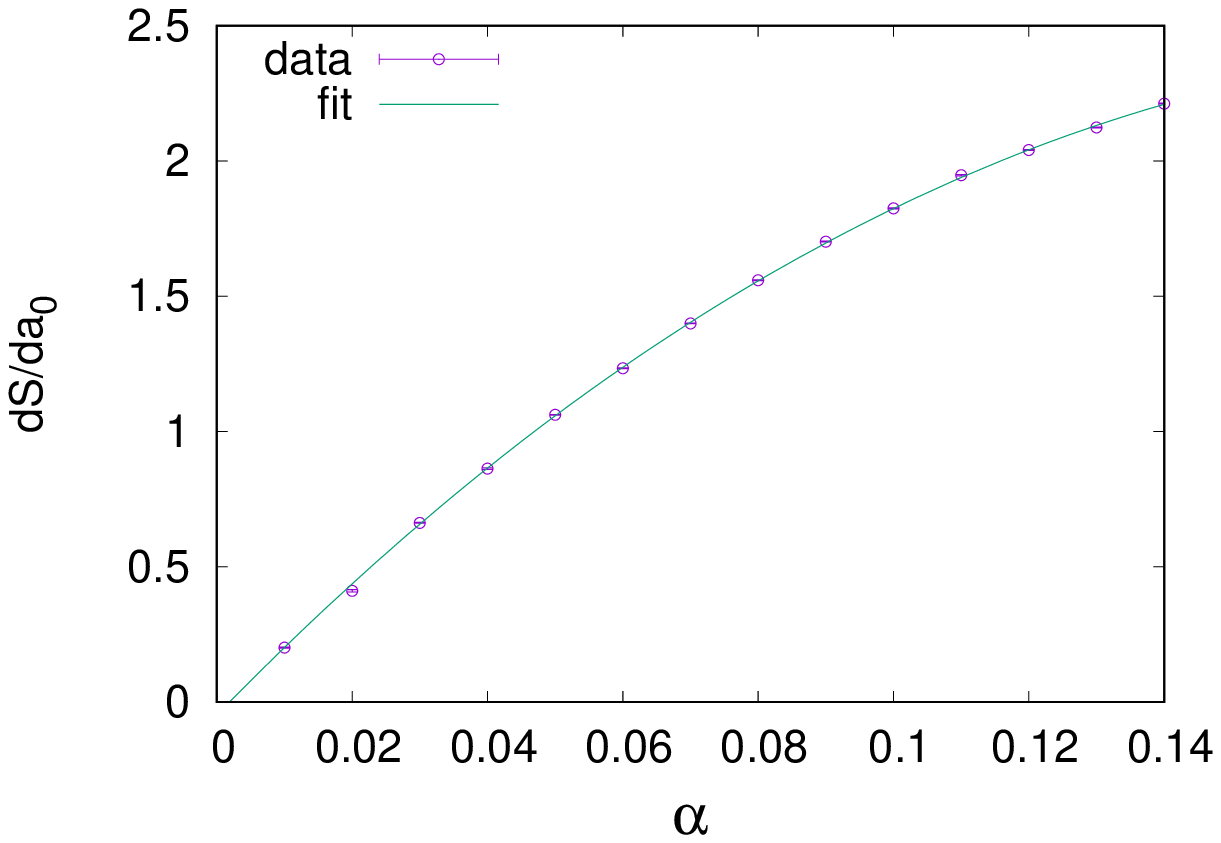}
\label{dSb}
}
\caption{Data for $(dS/da_0)_\a$ vs.\ $\a$ at (a) $\b=5.7, ma=0.6$, where $\langle P \rangle=0.012$, and at (b)
$\b=5.85, ma=0.435$, where $\langle P \rangle=0.11$.  Note the linearity of $dS/da_0$ in subfigure (a), and the non-linearity seen 
in subfigure (b), in the neighborhood of $\a=\langle P \rangle=0.11$. }
\label{dSk0}
\end{figure} 
    It is clear, since there should be only finite range correlations in the PLA, that $K(R)$ must die off faster than a power at large $R$, since
otherwise one could bring down terms in the effective action to produce a power-law falloff in the Polyakov line correlator.
However,  since $\tK(\vk)$ is determined at only a small set of $\vk$ values on a $16^3$ volume, it is necessary to fit the data to some analytic expression in order to carry out the inverse Fourier transform to $K(R)$.  It is of interest to see whether there are indications of the required
rapid falloff at large $R$.  As in our previous work  \cite{Greensite:2014isa,*Greensite:2013bya,Hollwieser:2016hne}, we fit the data for $\tK(\vk)$ at $\vk \ne 0$ by a double straight-line fit 
\bea
            \tK^{fit}(\vk) = \left\{ \begin{array}{cl}
                   c_1 - c_2 k_L & k_L \le k_0 \cr
                   d_1 - d_2 k_L & k_L \ge k_0 \end{array} \right. \ ,
\label{Kfit1}
\eea 
where 
\beq
           k_L = 2 \sqrt{ \sum_{i=1}^3 \sin^2(k_i/2)} 
\eeq
is the lattice momentum, and carry out the inverse Fourier transform, but taking $\tK(0)$ from the data rather than the fit.  The double-line fit  at $\b=5.7,~ma=0.6,~N_t=6$ is shown in Fig.\ \ref{Kfit}, and the corresponding inverse Fourier transform to $K(R)$ is displayed in Fig.\ \ref{KRall}.  Qualitatively, the behavior shown is typical of the data for $K(R)$ in all
of our simulations.  What we observe is that $K(R)$ falls off roughly as $1/R^4$ up to $R \approx 4$, as seen from a straight-line fit on the log-log plot of Fig.\ \ref{KR}.  Beyond $R=4$ $K(R)$ passes through zero (Fig.\ \ref{KRsmall}), after which the data oscillates and is rather noisy. We believe that the
noisy behavior is an artifact of having a small lattice, and a relatively small number of data points for $\tK(\vk)$.  But we do see, up to the
onset of noise, precisely the rapid drop to zero which is expected on general grounds.  We therefore discard the values of $K(R)$ beyond
$R=R_{cut}$, and reset those values to 
\beq
         K(R>R_{cut}) = 0
\eeq
where $R_{cut}$ is taken as the point where $K(R)$ is either zero, or else reaches a minimum near zero before oscillating.

   We have also tried to fit the data for $\tK(k)$ to a third-order polynomial.  Although the resulting $K(R)$ agrees with our double-line
method up to $R \approx 4$, it then goes asymptotically to a non-zero constant, which is not the right behavior.  We therefore prefer the original double line fit used in our previous work, which at least indicates that $K(R)$ should be negligible beyond some $R_{cut}$, and
which also gives us a fairly precise criterion for the choice of $R_{cut}$.  In fact by this criterion we find $R_{cut}=4.6$ in all cases.  We report on the effects of small variations in the choice of
$R_{cut}$ below in section \ref{results}.

   With $K(R)$ determined by the procedure just outlined, we find $h$ by simulating the PLA in eq.\ \rf{eq:SP}  at $\m=0$ with a series of trial values of the $h$ parameter, until $\langle P \rangle$ computed in the PLA agrees with the value computed in the LGT up to error bars.  This is not entirely straightforward.  As we found in our earlier work \cite{Hollwieser:2016hne}, the highly non-local term containing
${K(\vx-\vy)}$ in the action leads to metastable states which, in a lattice simulation of the PLA, depend on the initialization and persist for thousands of iterations.  A cold start, which sets $P(\vx)=1$ initially, generally leads to the system in the ``deconfined'' phase, with a large expectation value $\langle P \rangle$, even at $h=0$.  This is in strong disagreement with the LGT.  
Instead we initialize the system at $P(\vx)=0$, which then stays in the ``confined'' phase.

   In principle $h$ could be determined from our data via eq.\ \rf{K0a}, or the simplified version in eq.\ \rf{deriv1}.  We have not followed
this approach for two reasons.  First, the value of $h$ turns out to be very small at $\b=5.7$ and below, and the value determined
from a straight-line fit of data for the left hand side of \rf{deriv1} vs.\ $\alpha$ has a very large error bar.  As an example, we plot this data,
and the corresponding straight-line fit, in Fig.\ \ref{dSa} for the case of $\b=5.7$.  In this case a straight-line fit gives $h=0.0056 \pm 0.0023$, where the error bar is about 40\% of the estimate.  The alternative procedure used in this paper, of choosing $h$ to get agreement for the Polyakov line value in the PLA and the LGT at $\m=0$, gives $h=0.0042$, which is actually well within the sizeable error bar associated with the straight-line fit procedure. But the relative error $\D h/h$ only gets worse at smaller $\b$, and for this reason we have resorted to the Polyakov-fit method to determine $h$.

\begin{figure*}[htb]
\subfigure[~$\b=5.63,~ma=0.711$]  
{   
 \includegraphics[scale=0.6]{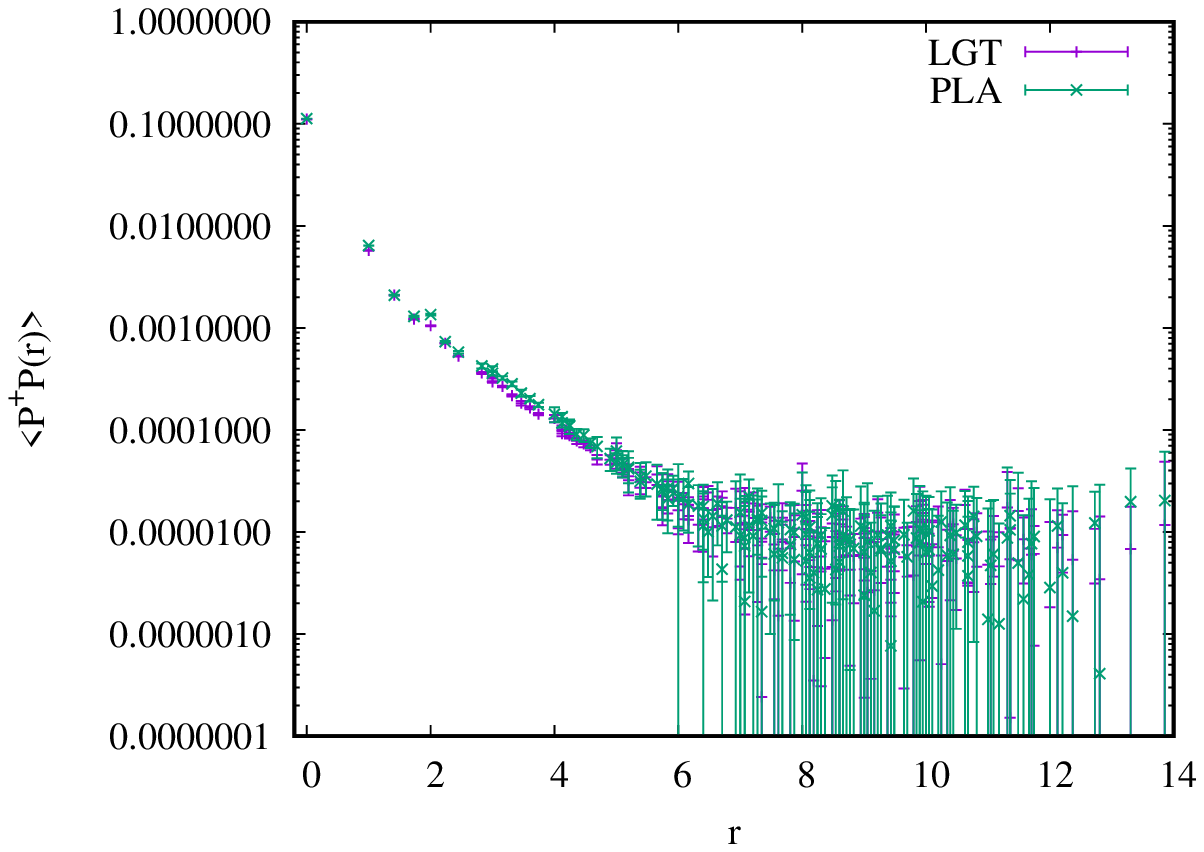}
}
\subfigure[~$\b=5.70,~ma=0.601$]  
{   
 \includegraphics[scale=0.6]{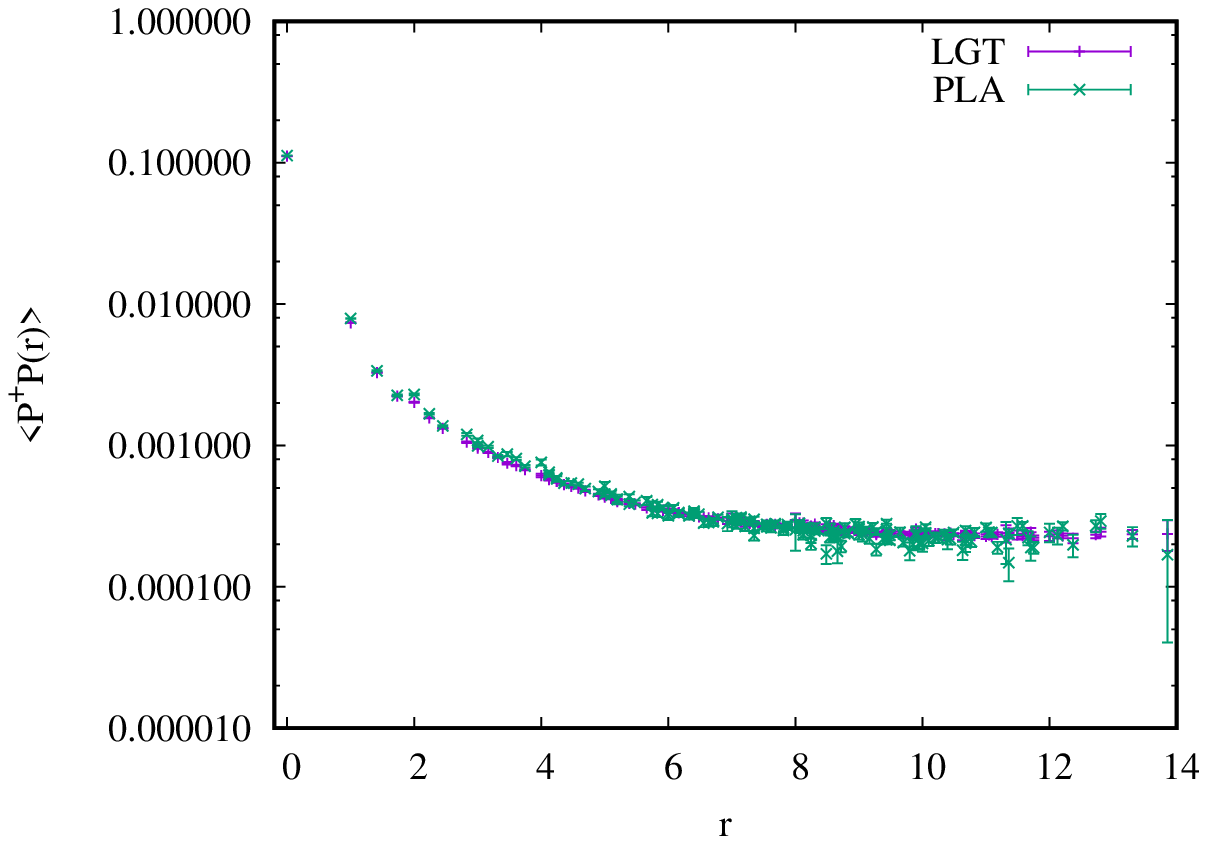}
}
\subfigure[~$\b=5.75,~ma=0.536$]  
{   
 \includegraphics[scale=0.6]{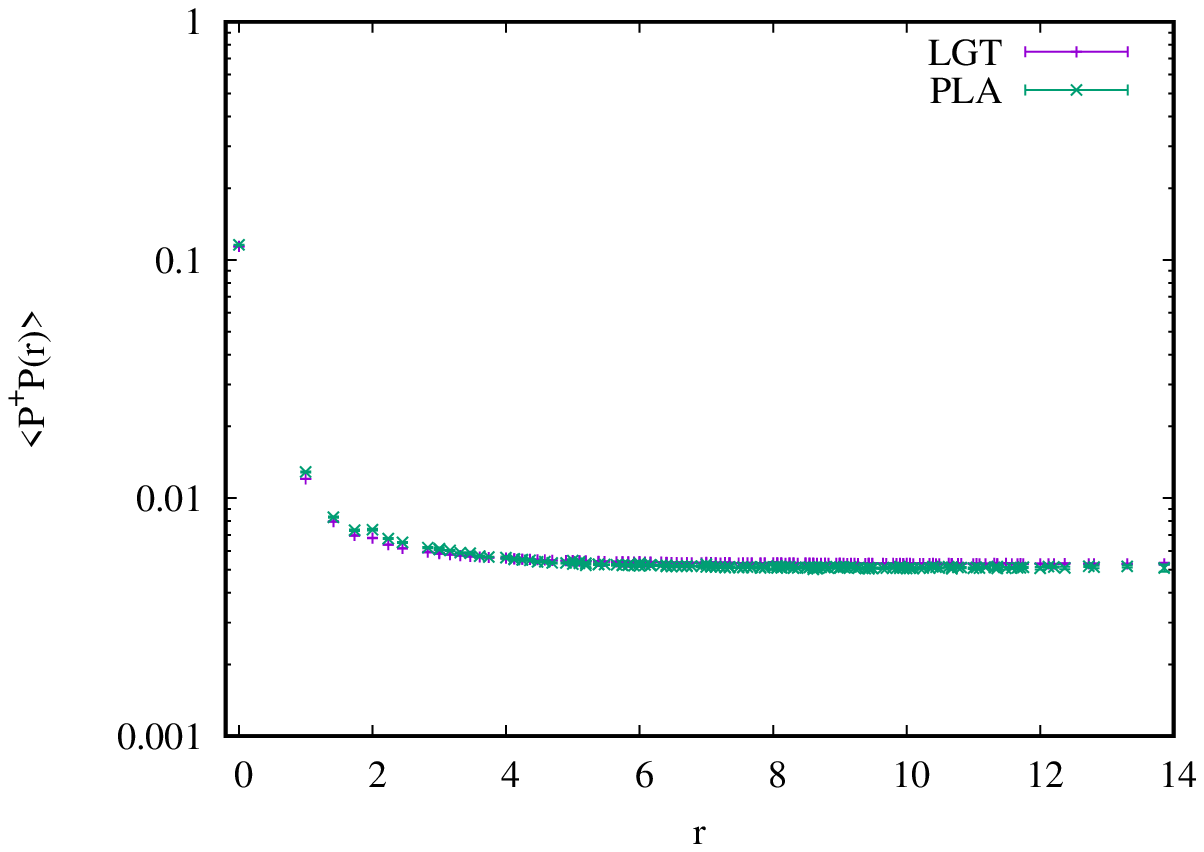}
}
\subfigure[~$\b=5.80,~ma=0.482$]  
{   
 \includegraphics[scale=0.6]{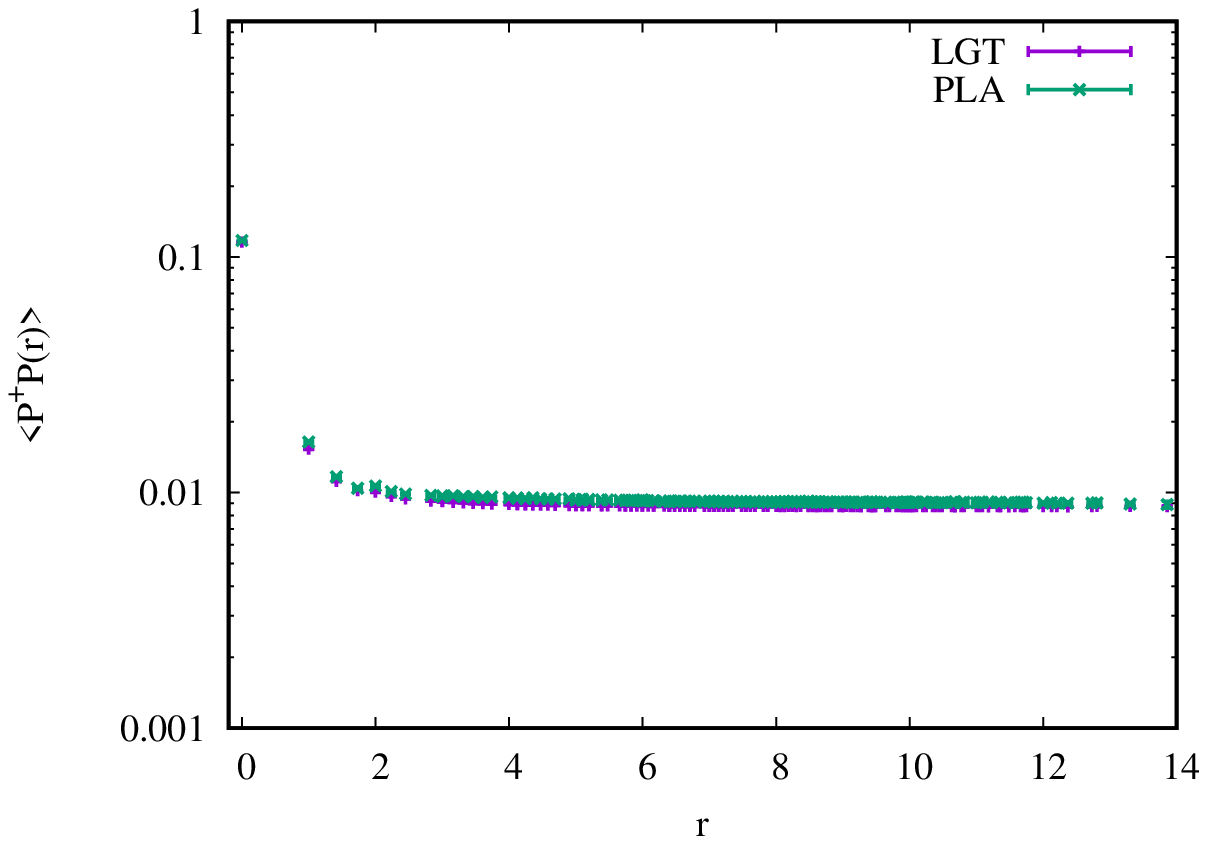}
}
\caption{Comparison of Polyakov line correlators $\langle P(0) P^\dg(r)$ vs.\ $r$ at $\m=0$ and $m=$695 MeV obtained from simulations of the effective PLA and underlying LGT.  The LGT lattice volume is $16^3 \times 6$, and different temperatures are obtained by varying the lattice coupling:
(a) $T=163$ MeV, (b) $T=193$ MeV, (c) $T=216$ MeV, (d) $T=241$ MeV.}
\label{plcors}
\end{figure*}

    $h$ rises rapidly above $\b=5.7$, but here we encounter a different kind of problem: the data for $dS/da_0$ as a function of
$\a$ does not fit a straight line, particularly for $\a$ at the Polyakov line expectation value $\langle P \rangle$.   
This is illustrated in Fig.\ \ref{dSb}.  One could try to fit the right hand side of \rf{deriv1} to the tangent at $\a= \langle P \rangle$, but we believe it is better to use a consistent procedure for all $\b$ values, so we have derived $h$ above and below $\b=5.7$ in the same way, choosing an $h$ to get $\langle P \rangle$ in agreement with the lattice gauge theory value.

Having obtained $h$, we can then go on to compute the Polyakov line correlator at ${\m=0}$
\beq
          G(|\vx-\vy|) = \langle P(\vx) P(\vy) \rangle \ ,
\label{Pcorr}
\eeq
in the PLA, and compare with the corresponding correlator obtained in the underlying LGT.   The results are in very good agreement; examples are
shown in Fig.\ \ref{plcors} for $\beta=5.63,5.7,5.75,5.8$, corresponding to temperatures $T=163,193, 216, 241$ MeV respectively.    

\section{Mean field at $\m \ne 0$}

    The PLA still has a sign problem at $\m \ne 0$, which we deal with via a mean field method \cite{Greensite:2012xv,*Greensite:2014cxa}.  We summarize here only the essential points; a detailed derivation may be found in the cited references.

    Starting from the partition function of the effective theory
\bea
   Z &=& \int \prod_\vx dU_\vx \Dt_\vx(\m,\tr U,\tr U^\dg) e^{S_0}
\non \\
   S_0 &=& \sum {1\over 9} K(\vx-\vy) \tr U_\vx \tr U_\vy \ ,
\eea
where $\Dt_\vx(\m,\tr U,\tr U^\dg)$ is the product of the determinant factors \rf{det1} and \rf{det2}, we rewrite
\bea
S_0 &=& J_0 \sum_\vx (v \tr U_\vx + u \tr U_\vx^\dg) - uvJ_0V 
\non \\
& &+ a_0 \sum_{\vx} \tr[U_\vx] \tr[U_\vx^\dg]+E_0 \ ,
\eea
where $V=L^3$ is the lattice volume, and we have defined
\bea
     E_0 &=& \sum_{(\vx \vy)} (\tr U_\vx-u)(\tr U_\vy^\dg - v) \on K(\vx-\vy) \ ,
\non \\
     J_0 &=&  {1\over 9} \sum_{\vx \ne 0} K(\vx)  ~~~,~~~ a_0 = {1\over 9} K(0) \ .
\eea
Parameters $u$ and $v$ are to be chosen such that $E_0$ can be treated as a perturbation, to be ignored as a first approximation. In particular, $\langle E_0 \rangle = 0$ when
\bea
u = \langle \tr U_x \rangle ~~~,~~~ v = \langle \tr U^\dg_x \rangle \ ,
\label{consistency}
\eea
and these conditions turn out to be equivalent to the stationarity of the mean field free energy.  The mean field approximation is obtained,
at leading order, by dropping $E_0$, in which case the partition function factorizes, and can be solved analytically as a function
of $u$ and $v$.  After some manipulations (cf.\ \cite{Greensite:2012xv,*Greensite:2014cxa}), one finds the mean field approximations
$u, v$ to $\langle \tr U_x \rangle$ and $\langle \tr U^\dg_x \rangle$ respectively, by solving the pair of equations
\bea
u - {1\over G}{\pa G \over \pa A}=0  ~~~~~\text{and}~~~~~ v - {1\over G}{\pa G \over \pa B}=0 \ ,
\label{hq-conditions}
\eea
where $A=J_0 v, ~ B=J_0 u$.  The expression $G(A,B)$ is given by
\bea
G(A,B) =  \Dt\left(\m,{\pa \over \pa A},{\pa \over \pa B} \right) 
 \sum_{s=-\infty}^{\infty}   \det\Bigl[D^{-s}_{ij} I_0[2\sqrt{A B}] \Big]  \ ,
\label{G}
\eea
where $D^{-s}_{ij}$ is the $i,j$-th component of a matrix of differential operators
\bea
D^s_{ij} &=& \left\{ \begin{array}{cl}
                         D_{i,j+s} & s \ge 0 \cr
                         D_{i+|s|,j} & s < 0 \end{array} \right. \ ,
\non \\
D_{ij} &=& \left\{ \begin{array}{cl}
                         \left({\pa \over \pa B} \right)^{i-j} & i \ge j \cr 
                        \left({\pa \over \pa A} \right)^{j-i} & i < j \end{array} \right. \ ,
\eea
The mean field free energy density $f_{mf}$ and fermion number density $n$ are
\bea
           {f_{mf} \over T} &=& J_0 u v - \log G(A,B) \non \\
           n &=& {1\over G} {\pa G \over \pa \m} \ .
\eea

   The stationarity conditions \rf{hq-conditions} may have more than one solution, and here it is important to take account of
the existence of very long lived metastable states in the PLA.  The state at $\m=0$ which corresponds to the LGT is the one
obtained by initializing at $P_\vx=0$, and in the mean field analysis this is actually not the state of lowest free energy (its stability
in a Monte Carlo simulation is no doubt related to the highly non-local couplings in the PLA).  By analogy, 
at finite $\m$ we look for solutions of \rf{hq-conditions} by starting the search at $u=v=0$, regardless of whether another solution exists at a slightly lower $f_{mf}$.
  
     The mean field calculation at any chemical potential $\m$ requires three numbers: $a_0$ and $J_0$, which are both derived from $K(\vx)$, and $h$.  Denote by h(mfd) the value of $h$ for which mean field gives a Polyakov line $\langle P \rangle$, at $\m=0$, in agreement with lattice gauge theory.  This can be compared to the value of $h$, denoted h(PLA), for which the Polyakov line computed in the PLA  at $\m=0$ agrees with the lattice gauge theory value.  The values of h(mfd) and h(PLA) at the points we have computed can be compared in Table \ref{table1}; in general they are not far off, which is some evidence of the validity of the mean field approach in this application.
  
\section{\label{results} Results}

    Our LGT numerical simulations were carried out in SU(3) lattice gauge theory with unrooted staggered fermions on a $16^3 \times 6$ lattice.   For scale setting we have taken the lattice spacing from the Necco-Sommer expression \cite{Necco:2001xg}
\bea
a(\beta) &=& (0.5 ~ \mbox{fm}) \exp\left[-1.6804 - 1.7331(\beta-6)  \right. \non \\
& &  \left. + 0.7849(\beta-6)^2 - 0.4428(\beta-6)^3 \right] \ .
\eea
We take the quark mass in lattice units to be $ma=0.6$ at $\beta=5.7$.  This corresponds to a mass of $m=695$ MeV, and temperature 
$T=193$ MeV in physical units.  We keep the physical mass and the extension $N_t=6$ in the time direction fixed, and vary the 
temperature by varying the lattice spacing, i.e.\ by varying $\beta$.

\begin{figure*}[htb]
\subfigure[~$\b=5.60,~ma=0.767$]  
{   
 \includegraphics[scale=0.6]{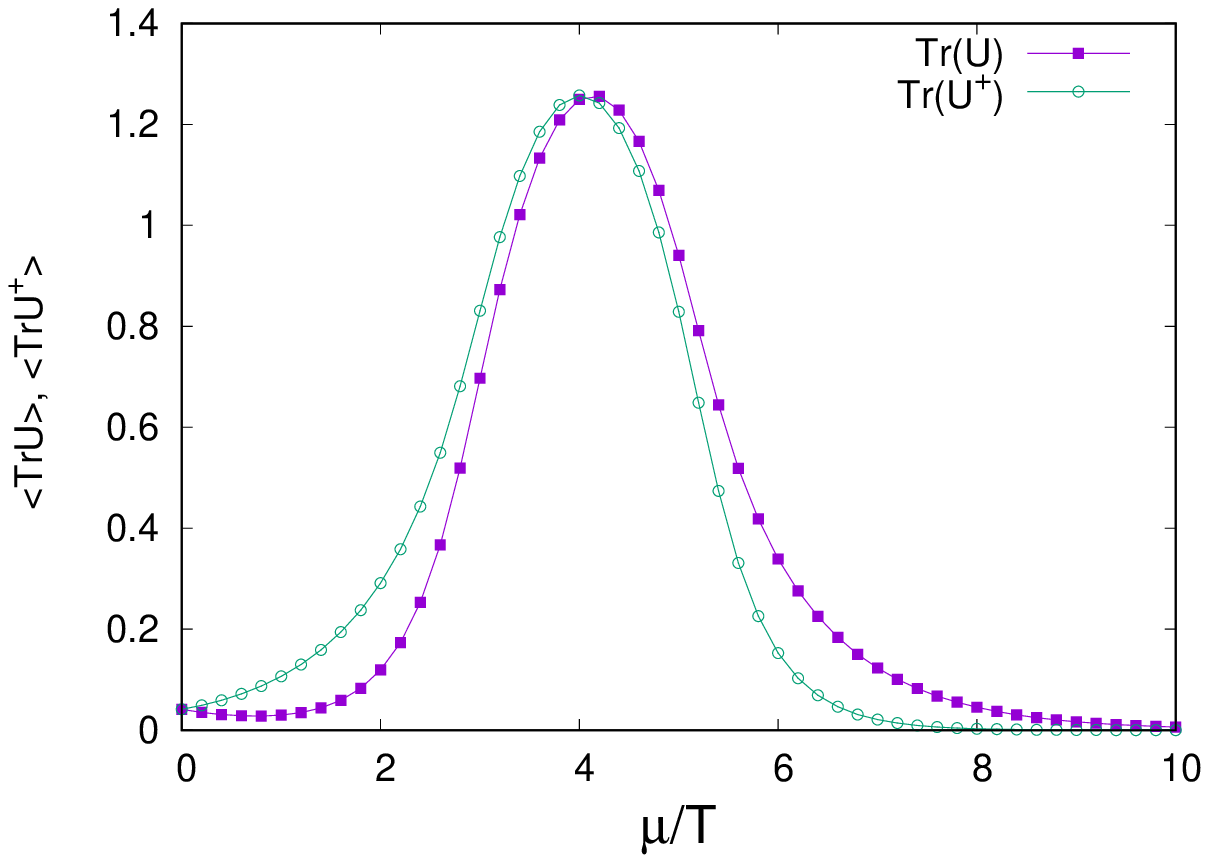}
}
\subfigure[~$\b=5.70,~ma=0.6$]  
{   
 \includegraphics[scale=0.6]{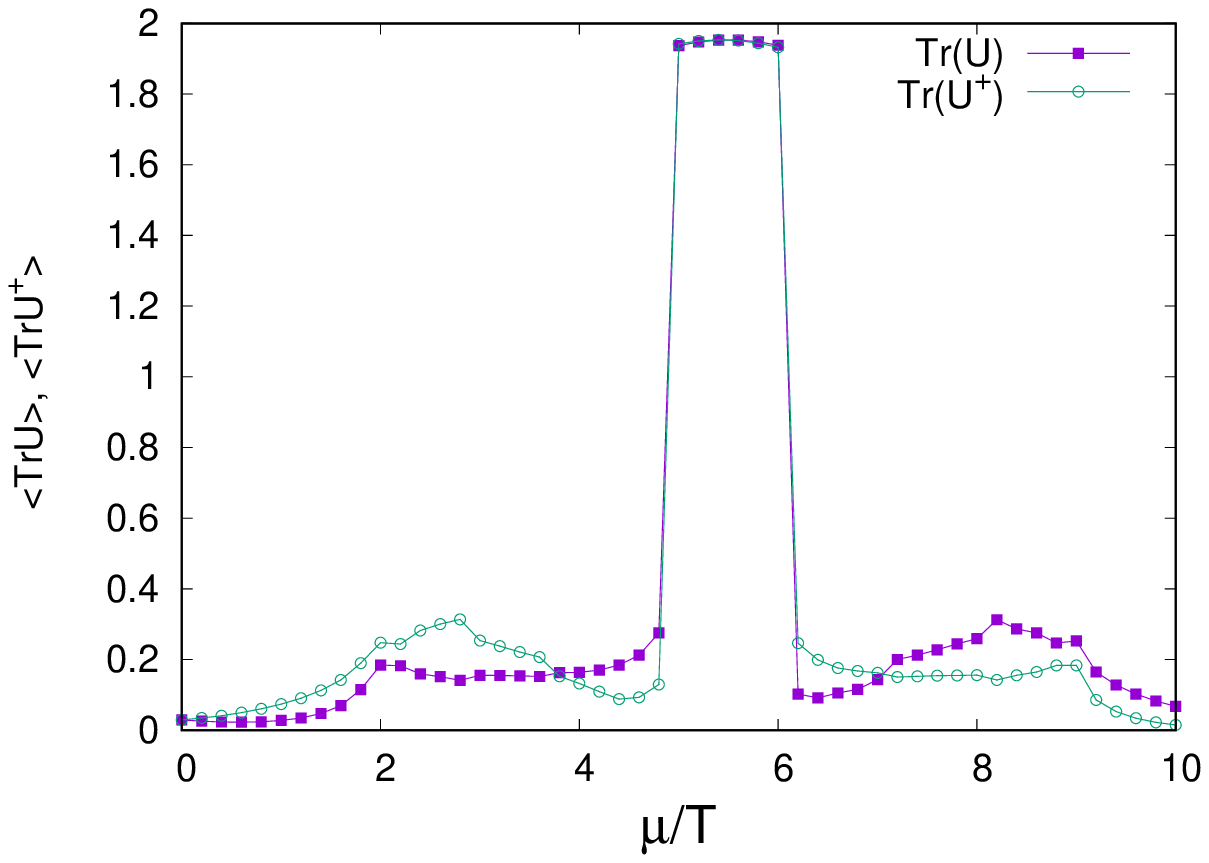}
}
\subfigure[~$\b=5.75,~ma=0.536$]  
{   
 \includegraphics[scale=0.6]{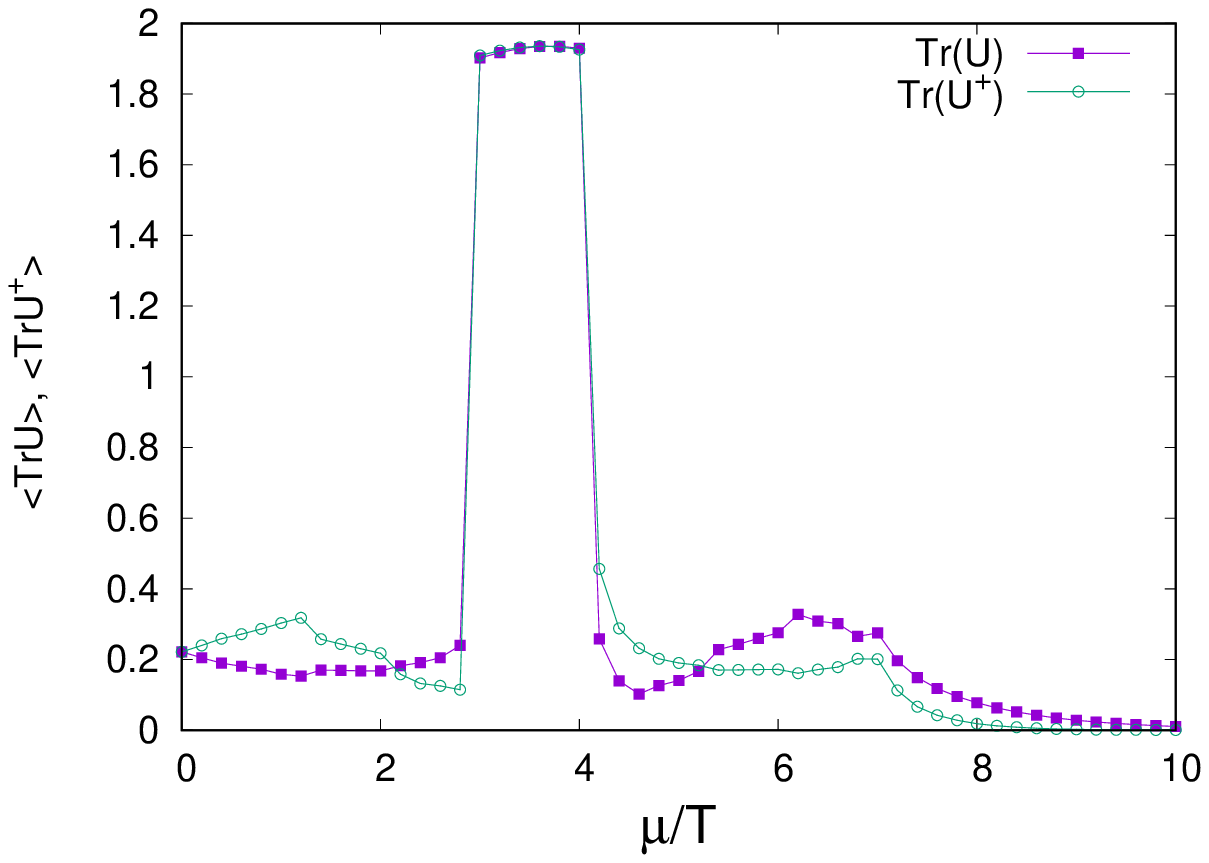}
}
\subfigure[~$\b=5.85,~ma=0.435$]  
{   
 \includegraphics[scale=0.6]{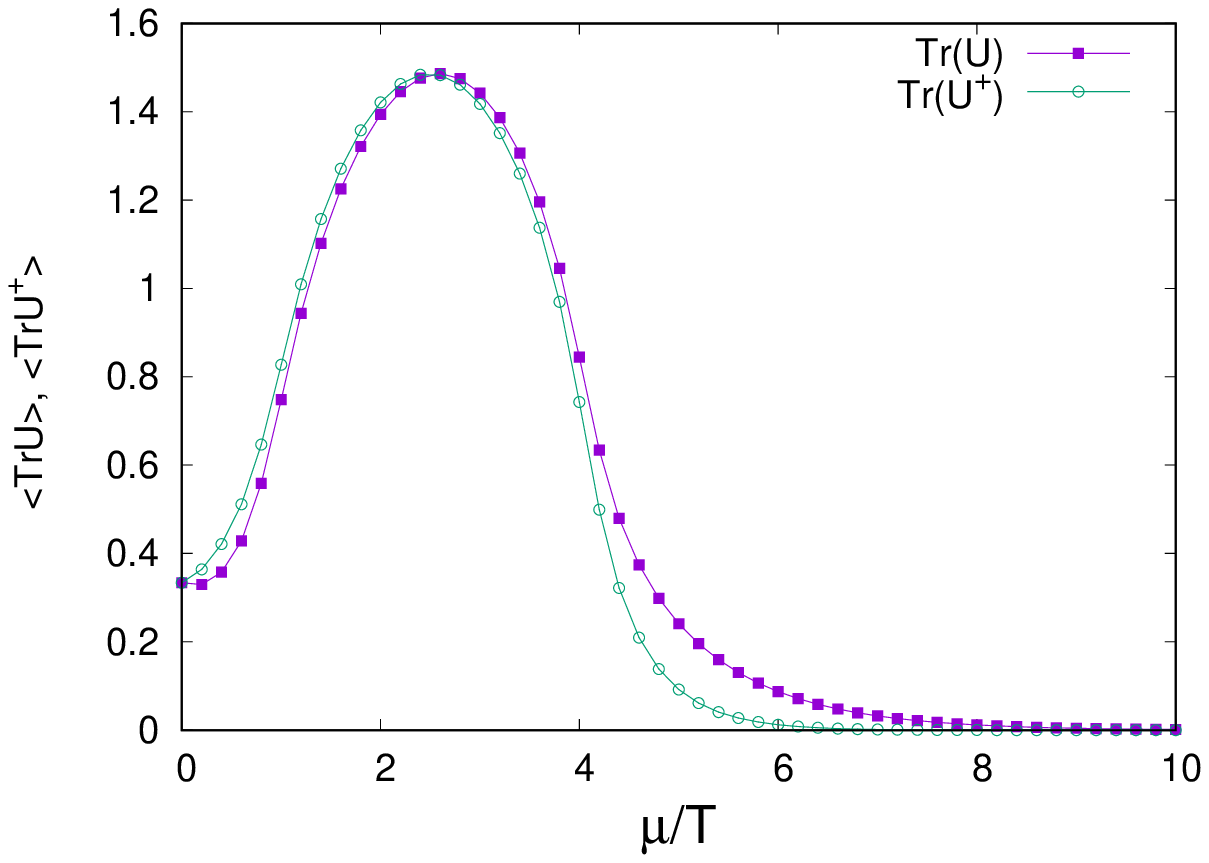}
}
\caption{Examples of mean-field calculations of $\langle \tr U \rangle$ and $\langle \tr U^\dg \rangle$ at finite $\m$ at: (a) $T=151$ MeV; (b) $T=193$ MeV; (c) $T=216$ MeV; (d) $T=267$ MeV. There are phase
transitions in the two mid-range temperatures, but no transitions at the highest and lowest temperatures.}
\label{uv}
\end{figure*}

\begin{figure}[htb]
 \includegraphics[scale=0.6]{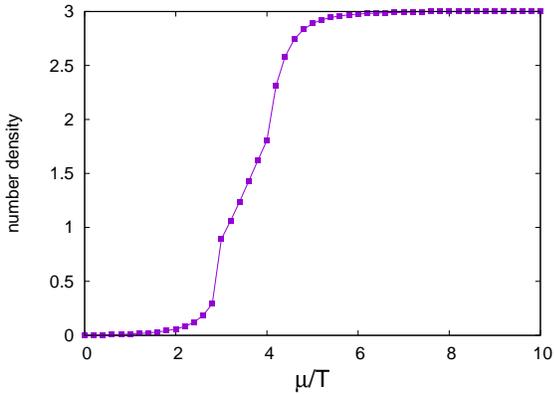}
\caption{Mean field result for fermion number density per lattice site, vs.\ $\m/T$, at $\b=5.75$ corresponding to $T=216$ MeV.  Two transitions are visible.  Note that the transition at higher chemical potential occurs close to the saturation value, which is  $n=3$ for staggered fermions.
Since saturation is a lattice artifact, this second transition is probably unphysical.}
\label{density}
\end{figure}

    Given the PLA at $\m=0$, the Polyakov line expectation values $\langle \tr U_\vx \rangle$ and $\langle \tr U^\dg_\vx \rangle$ at
finite $\m$ are calculated by the mean field method outlined above, with a sample of our results displayed in Fig.\ \ref{uv}.  A discontinuity in a plot of $\langle \tr U_\vx \rangle, \langle \tr U^\dg_\vx \rangle$ vs.\ $\m$ is the sign of a transition at finite density, and
conversely the absence of any discontinuity indicates the absence of any transition.  When a transition occurs at some value 
of chemical potential $\m_1$, then there is a second transition at some $\m_2 > \m_1$.   However, while the first transition occurs
at some relatively low density (in lattice units) on the order of $n \approx 0.2$, the second transition always occurs at a density
close to the saturation value, which for staggered fermions is $n=3$.  An example of density $n$ vs.\ $\m$ at $\b=5.75$ is shown in 
Fig.\ \ref{density}; transitions occur at the sharp jumps in density. Since the saturation value is a lattice artifact, we do not attach much physical significance to the second transition at $\m=\m_2$.  Transition points listed in our tables and plots all correspond to $\m=\m_1$.

   Our full set of results, including the parameters used in the mean field calculation, is shown in Table \ref{table1}.  As a check on the mean field calculation we have calculated $h$ in two ways: 
\begin{itemize} 
\item by choosing $h$ such that the mean field calculation at $\m=0$ reproduces  the Polyakov line
value of lattice gauge theory. These values are denoted h(mfd) in Table \ref{table1}.  The corresponding first transition $\m_1/T$,
and $\m_1$ in MeV, are displayed in the columns adjacent to h(mfd). 
\item by choosing $h$ such that a simulation of the PLA at $\m=0$ reproduces the Polyakov line
value of lattice gauge theory, as described in section III.  These values are denoted h(PLA) in Table \ref{table1}.  The corresponding first transition $\m_1/T$, and $\m_1$ in MeV, are displayed in the columns adjacent to h(PLA). 
\end{itemize}
The fairly close agreement between h(mfd) and h(PLA), in what amounts to a spin system in $D=3$ dimensions, can be attributed to the highly non-local nature of the PLA.  Mean field treatments work best when each
degree of freedom is coupled to many other degrees of freedom.  For nearest-neighbor couplings, this generally means that the treatment is only accurate in higher dimensions.  But it seems that the relatively long range of $K(R)$, which results in each Polyakov line being 
coupled to many other lines, serves the same purpose as high dimensions in a nearest-neighbor theory.  We should also note the good agreement found in ref.\ \cite{Greensite:2014cxa} between Langevin simulations at finite density in a spin system of this type, with the mean field treatment we have discussed.

\begin{table*}
\begin{tabularx}{1.0\textwidth}{*{14}{|Y}|}
\hline
		$\beta$&$T$[MeV]&$a$[fm]&$ma$&$\langle P\rangle$&$K(0)$&$a_0$&$J_0$&$h$(mfd)&$\mu/T$&$\mu$[MeV]&$h$(PLA)&$\mu/T$&$\mu$[MeV]\\
\hline
5.60 & 151 & 0.217 & 0.767 & 0.00135 & 6.729 & -0.0113 & 0.5753 & 0.00176 & - & - & 0.00170 & - & - \\
5.63 & 163 & 0.201 & 0.711 & 0.00188 & 7.648 & -0.0079 & 0.6803 & 0.00183 & - & - & 0.00176 & - & - \\
5.64 & 167 & 0.196 & 0.694 & 0.00218 & 7.923 & -0.0121 & 0.7117 & 0.00191 & - & - & 0.00183 & - & - \\
5.65 & 171 & 0.192 & 0.677 & 0.00254 & 8.238 & -0.0190 & 0.7795 & 0.00180 & 4.775 & 817 & 0.00171 & 4.825 & 825 \\
5.66 & 176 & 0.187 & 0.660 & 0.00318 & 8.764 & -0.0201 & 0.8171 & 0.00200 & 5.275 & 928 & 0.00183 & 4.825 & 849 \\
5.68 & 184 & 0.178 & 0.630 & 0.00558 & 9.069 & -0.0192 & 0.8381 & 0.00288 & 5.075 & 934 & 0.00252 & 5.225 & 961 \\
5.70 & 193 & 0.170 & 0.601 & 0.01198 & 9.382 & -0.0256 & 0.8646 & 0.00513 & 4.625 & 893 & 0.00424 & 4.875 & 941 \\
5.73 & 206 & 0.159 & 0.561 & 0.05734 & 10.221 & -0.0360 & 0.8709 & 0.01527 & 3.525 & 726 & 0.01353 & 3.675 & 757 \\
5.75 & 216 & 0.152 & 0.536 & 0.07235 & 9.851 & -0.0334 & 0.8608 & 0.02971 & 2.825 & 610 & 0.02543 & 2.975 & 643 \\
5.77 & 226 & 0.145 & 0.513 & 0.08354 & 9.760 & -0.0380 & 0.7753 & 0.03940 & 1.125 & 254 & 0.03763 & 1.175 & 266 \\
5.775 & 229 & 0.144 & 0.508 & 0.08522 & 9.719 & -0.0364 & 0.7920 & 0.03530 & 1.825 & 418 & 0.03333 & 1.875 & 429 \\
5.78 & 231 & 0.142 & 0.502 & 0.08703 & 9.834 & -0.0454 & 0.7622 & 0.04515 & 0.775 & 179 & 0.04383 & 0.825 & 191 \\
5.80 & 241 & 0.136 & 0.482 & 0.09332 & 10.039 & -0.0438 & 0.7623 & 0.04639 & 0.675 & 163 & 0.04567 & 0.725 & 175 \\
5.85 & 267 & 0.123 & 0.435 & 0.10992 & 10.151 & -0.0540 & 0.6850 & 0.07716 & - & - & 0.07766 & - & - \\ 
\hline
\end{tabularx}
\caption{Simulation parameters for the LGT (Wilson gauge action and dynamical staggered fermions with $m_q=695$MeV on $16^3$x$6$ lattices), together with parameters for the corresponding  mean field computations and the resulting transition points. $R_{cut}=4.6$ in all cases. The $h$ values are chosen to reproduce the correct LGT value of $\langle P\rangle$ at $\m=0$, computed either by mean field (h(mfd)) or by simulation of the PLA (h(PLA)).  The transition points derived from each choice of $h$ are displayed in the columns to the right of h(mfd) and h(PLA) respectively.} 
\label{table1}
\end{table*}

\begin{table*}
\begin{tabularx}{1.0\textwidth}{*{6}{|Y}|*{4}{|Y}|*{4}{|Y}|}
\hline
\multicolumn{6}{|c||}{}&\multicolumn{4}{c||}{$R_{cut}=4.4$}&\multicolumn{4}{c|}{$R_{cut}=4.8$}\\
\hline
$\beta$&$T$[MeV]&$a$[fm]&$ma$&$\langle P\rangle$&$K(0)$&$J_0$&$h$(mfd)&$\mu/T$&$\mu$[MeV]&$J_0$&$h$(mfd)&$\mu/T$&$\mu$[MeV]\\
\hline
5.60 & 151 & 0.217 & 0.767 & 0.00135 & 6.729 & 0.5743 & 0.00176 & - & - &
		0.5748 & 0.00176 & - & - \\
5.63 & 163 & 0.201 & 0.711 & 0.00188 & 7.648 & 0.6783 & 0.00184 & - & - &
		0.6796 & 0.00184 & - & - \\
5.65 & 171 & 0.192 & 0.677 & 0.00254 & 8.238 & 0.7728 & 0.00185 & 4.375 &
		748 & 0.7790 & 0.00180 & 5.225 & 893 \\
5.66 & 176 & 0.187 & 0.660 & 0.00318 & 8.764 & 0.8072 & 0.00210 & 4.625 &
		814 & 0.8182 & 0.00199 & 5.275 & 928 \\
5.68 & 184 & 0.178 & 0.630 & 0.00558 & 9.069 & 0.8289 & 0.00304 & 4.975 &
		915 & 0.8386 & 0.00288 & 5.125 & 943 \\
5.70 & 193 & 0.170 & 0.601 & 0.01198 & 9.382 & 0.8549 & 0.00548 & 4.525 &
		873 & 0.8651 & 0.00511 & 4.675 & 902 \\
5.73 & 206 & 0.159 & 0.561 & 0.05734 & 10.221 & 0.8588 & 0.01733 & 3.375 &
		695 & 0.8722 & 0.01505 & 3.575 & 736 \\
5.75 & 216 & 0.152 & 0.536 & 0.07235 & 9.851 & 0.8463 & 0.05950 & 2.075 &
		448 & 0.8637 & 0.02423 & 3.075 & 664 \\
5.77 & 226 & 0.145 & 0.513 & 0.08354 & 9.760 & 0.7646 & 0.04198 & 1.075 &
		243 & 0.7766 & 0.03908 & 1.175 & 266 \\
5.775 & 229 & 0.144 & 0.508 & 0.08522 & 9.719 & 0.7816 & 0.03789 & 1.325 &
		303 & 0.7929 & 0.03506 & 1.725 & 395 \\
5.78 & 231 & 0.142 & 0.502 & 0.08703 & 9.834 & 0.7591 & 0.04594 & 0.725 &
		167 & 0.7620 & 0.04521 & 0.775 & 179 \\
5.80 & 241 & 0.136 & 0.482 & 0.09332 & 10.039 & 0.7516 & 0.04929 & 0.725 &
		175 & 0.7638 & 0.04599 & 0.975 & 235 \\
5.85 & 267 & 0.123 & 0.435 & 0.10992 & 10.151 & 0.6767 & 0.07974 & - & - &
		0.6867 & 0.07663 & - & - \\ 
\hline\end{tabularx}
\caption{Sensitivity of the transition point to $\pm 0.2$ variations (in lattice units) in the value of $R_{cut}=4.6$, which was determined by the procedure described in the text.  In these calculations, h(mfd) is used throughout.}
\label{table2}
\end{table*}

\begin{figure*}[htb]
\subfigure[~analytic continuation]  
{   
 \includegraphics[scale=0.6]{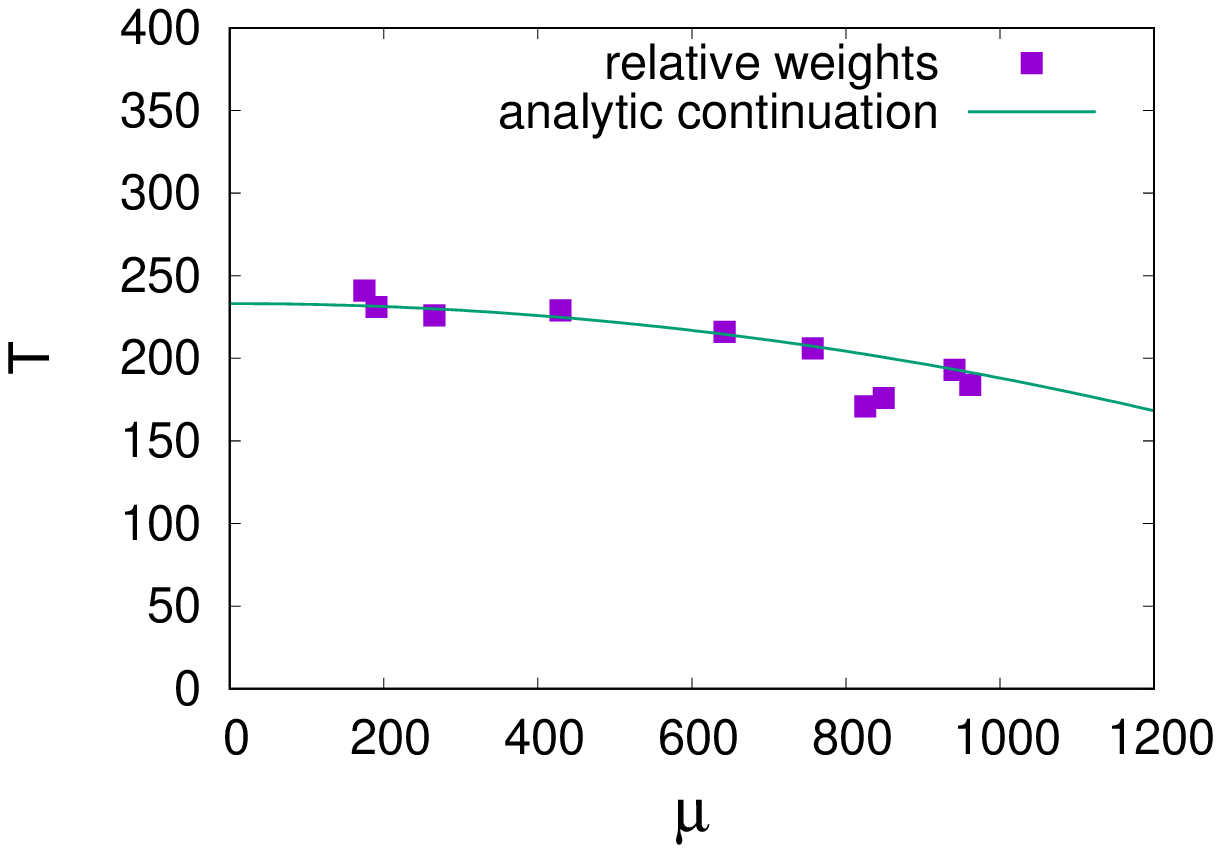}
\label{maria}
}
\subfigure[~CL heavy dense]  
{   
 \includegraphics[scale=0.6]{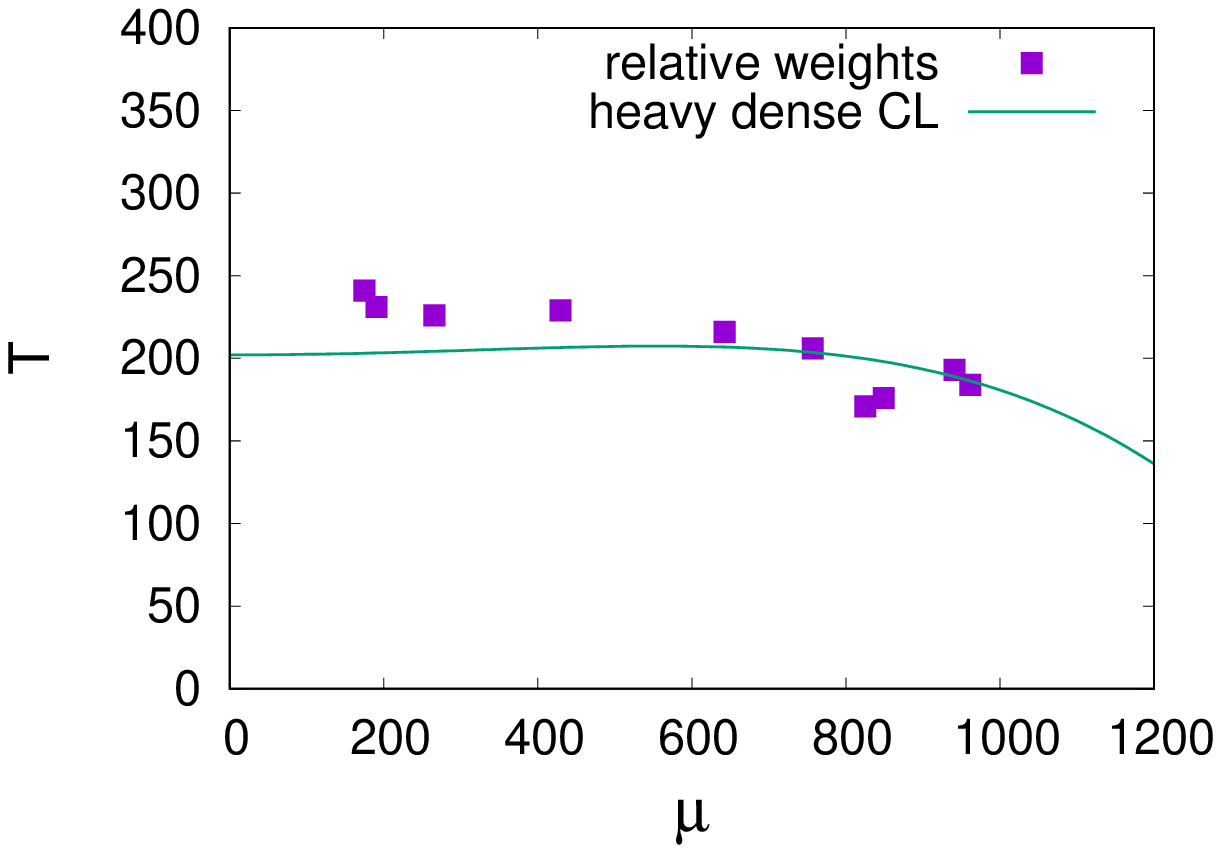}
\label{gert}
}
\caption{The phase transition line in the $\m-T$ plane for staggered unrooted fermions of mass
$m=695$ MeV, obtained by the method described in the text.  (a) Comparison with the analytic continuation formula of
d'Elia and Lombardo \cite{DElia:2002tig}. (b) Comparison with a complex Langevin expression for heavy dense quarks
\cite{Aarts:2016qrv}.}
\label{compare}
\end{figure*}

   Our results for the $\m,T$ transition points (using $h$=h(PLA)) for staggered, unrooted quarks of mass 695 MeV, are plotted in Fig.\ \ref{maria}. 
This figure is the main result of our paper, and holds for the temperature range $151 \le T \le 267$ MeV that we have
investigated.   We see that the phase transition line exists to an upper temperature of $T \approx 241$ MeV, where there is a critical
endpoint.  The fact that there is a critical endpoint at high temperature was expected.  What was unexpected is that the transition points
seem to disappear at lower temperatures.  The solid line is a comparison to the analytic continuation expression of d'Elia and Lombardo \cite{DElia:2002tig}, to be discussed further below.  

\begin{figure}[htb]
 \includegraphics[scale=0.6]{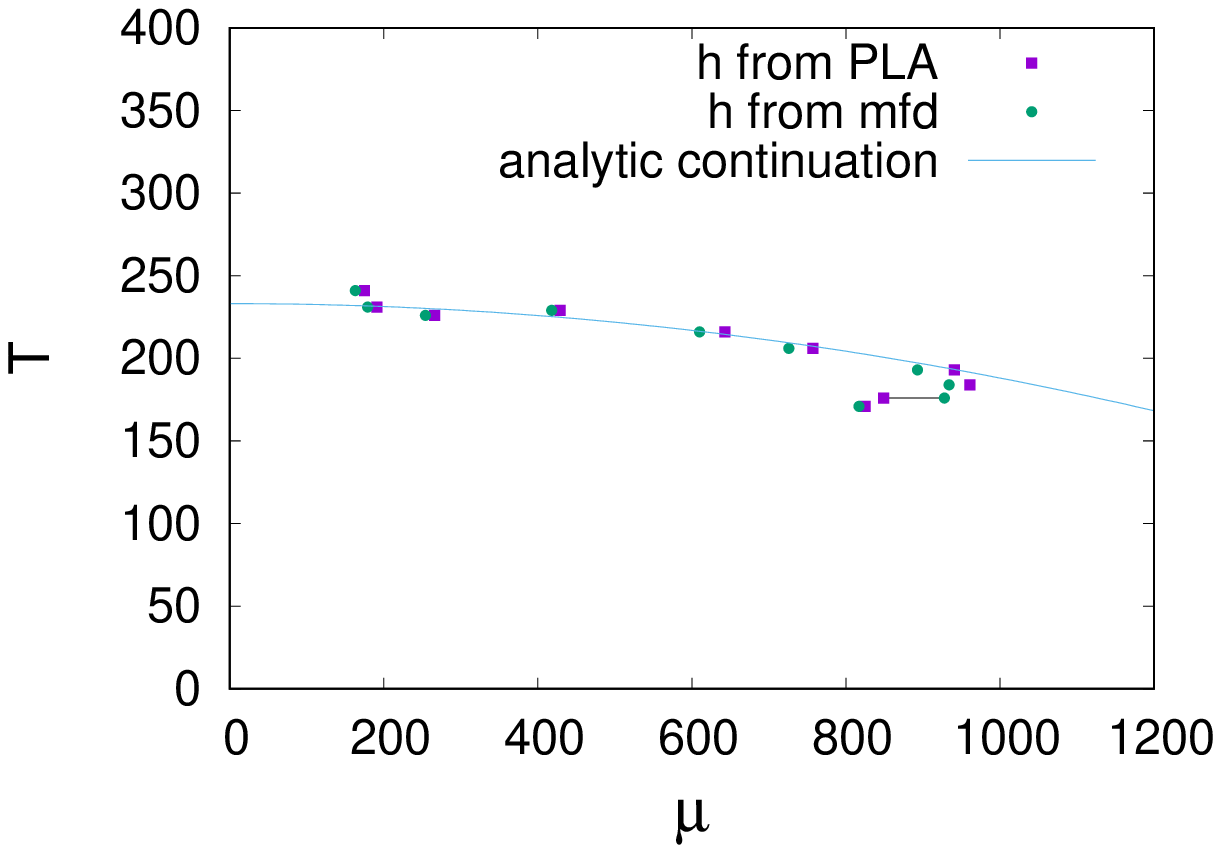}
\caption{Comparison of transition points derived using h(mfd) and h(PLA), as described in the text.  The corresponding
points are seen to be close to one another, with the exception of points obtained at $\b=5.66$, where the separation of corresponding transition points is indicated by a short horizontal line.}
\label{mfdPLA}
\end{figure}

    There are two transition points, computed at $\b=5.65$ and $\b=5.66$, which appear to be outliers on this plot, in the sense that they lie quite far from the analytic continuation curve.  These $\b$ values are in fact the smallest couplings at which we still see a transition, and differ sharply from the transition point at the next lowest $\b$ value, at $\b=5.68$ with the  transition at
$(\m,T)=(961,184)$ MeV.  We suspect that these two points are indicative of some instability in the mean field calculation in the immediate neighborhood of the $\b$ value where the phase transitions disappear, and we are inclined to discount them.  Taken literally, they would suggest that the transition line moves to lower values of $\mu$ as $T$ decreases below 184 MeV, which seems unlikely.  One indication
that something may be going wrong at these lowest $\b$ values is the fact that at $\b=5.66$, unlike at all our other data points, there is a very substantial disagreement between the transition point at $\m=928$, which is derived using h(mfd), and the transition point $\m=849$ MeV,
derived using h(PLA).    A comparison of transition points derived in the mean field approach using h(mfd) and h(PLA)  is displayed in Fig.\ \ref{mfdPLA}.  In this figure we have drawn a short horizontal line connecting transition points at $\b=5.66$, to indicate their separation.

    A possible source of systematic error could lie in our procedure for choosing $R_{cut}$, and it is worthwhile to check the sensitivity of
our results to a small variation in that parameter.  This is shown in Table \ref{table2}, where we have varied $R_{cut}$ (in lattice units) by 
$\pm 0.2$.  These should not be regarded as error bars, exactly, since the variation $\pm 0.2$ is rather arbitrary, but only as some indication of the sensitivity of our results to the value of $R_{cut}$.

     In any case, our results do indicate that the transition line does not continue all the way to $T=0$, and seems to end at about $T=184, ~\m=961$ MeV.  We cannot rule out the possibility that a high density transition reappears at some temperature lower than the lowest temperature (151 MeV) that we have considered.  Or perhaps the second critical endpoint goes away for light quarks. These possibilities we reserve for later investigation.

\subsection{Comparison with other results}

    There do exist results for the critical line in the $\m-T$ plane, expected to be valid at small $\m$, that have been obtained
by other methods.  Because these usually involve a choice of lattice fermions (Wilson, staggered), number of flavors, and quark masses different from our own, it is a little difficult to make a direct comparison with our data.  Nevertheless, it is interesting
make such comparisons anyway, for whatever they may be worth.

    Perhaps the most relevant comparison is to the analytic continuation method of d'Elia and Lombardo \cite{DElia:2002tig}, who, like us,
work with four flavors of staggered fermions.  Their approach is to find the transition line for imaginary chemical potential, and fit to a polynomial which is then analytically continued to real chemical potential.  The result (like the related Taylor expansion approach), is only
believable at small $\mu$.  However, these authors set $ma=0.05$, which means that their physical mass changes
at different $\b$.  What is found is that 
\beq
            T(\m) = T_c \left( 1 -  0.021{\m^2 \over 2T_c^2} \right)
\label{LD}
\eeq
where $T_c$ is the critical (or crossover) temperature at $\m=0$.  In order to make a comparison with our work we
just take $T_c$ as a fitting parameter, with the expectation that it can't be
too far from the quenched case of around $T_c=250$ MeV (in fact the fit comes out to be $T_c \approx 233$ MeV). 
The comparison of \rf{LD} with our result is shown in Fig.\ \ref{maria}, and there appears to be remarkably good agreement between the analytic continuation result, and the transition points we have found via relative weights, apart from the two outliers mentioned above.

        We have also made the comparison to the results for heavy dense quarks obtained by Aarts et al.\  \cite{Aarts:2016qrv} via the complex Langevin method.  That work employs Wilson fermions with two flavors, and a hopping parameter of $\kappa = 0.04$; i.e.\ extremely massive quarks.  These authors fit their data for the critical temperature at each $\m$ to a 2nd order
polynomial
\beq
            T(\m) = b_1(1-x) + b_2(1-x)^2
\label{Aa}
\eeq
where
\beq
              x = {\mu^2 \over \m_0^2}
\eeq
and $\m_0=-\ln(2\k)$.  This choice of $\mu_0$ is motivated by the hopping parameter expansion, and
the ``Silver Blaze'' phenomenon (nothing happens at $T=0$ until $\m=\m_0$).  Aarts et al.\  \cite{Aarts:2016qrv} report the
constants
\beq
              b_1 = 481  ~~~,~~~ b_2 = -279.3
\eeq
on their largest ($10^3$) lattice volume.  There is some volume dependence in these constants, and the result
is presumably only valid for small hopping parameter and large chemical potential.  In order to make some kind of comparison with this
with heavy-dense approach, we have just taken $\m_0$ to give the closest fit to our data points.  The result of this fit is shown in Fig.\
\ref{gert}.  It's hard to know how seriously to take this comparison, since Aarts et al.\ are using two flavors of Wilson fermions, as opposed to our four staggered flavors, and are working with extremely heavy quarks.   In any case, even allowing for a best fit value of $\m_0$, the heavy dense result seems quite far from our data, and perhaps this is not very surprising.

\section{Conclusions}

    We have found a first-order phase transition line for SU(3) gauge theory with dynamical unrooted staggered fermions of mass
695 MeV, by the method of relative weights combined with mean field theory, in the plane of chemical potential $\m$ 
and temperature $T$.   The critical line lies in a finite temperature range between $(\m,T)=(175,241)$ and $(\m,T)=(961,184)$ MeV and
the transition points lie, for the most part, along a line determined from analytic continuation of results at imaginary $\m$ \cite{DElia:2002tig}.

    We offer this result with reservations.  There is certainly a degree of arbitrariness in our approach, particularly in the choice of
ansatz for the PLA.  We have used a product of local determinants for the $\m$-dependent part of the PLA, and a non-local bilinear
form for the $\m$-independent part.  This form has given us an excellent match to the Polyakov line correlator at $\m=0$, computed
in the LGT.  But there is no guarantee that a more complicated form, involving e.g.\ Polyakov lines in higher representations, trilinear
couplings, etc., is not required at high densities.  It would be interesting to probe the existence and possible importance of such terms,
supplementing the method of relative weights and mean field with, e.g., the method of inverse Monte Carlo \cite{Wozar:2007tz,Bahrampour:2016qgw} and/or the approach of
Bergner et al.\ \cite{Bergner:2015rza}.  It would also be interesting to see whether the expected transition line reappears, for $m=695$ MeV, at temperatures below 151 MeV, or how the situation changes with lighter quark masses.  We reserve these questions for later study.

\acknowledgments{JG's research is supported by the U.S.\ Department of Energy under Grant No.\ DE-SC0013682.  RH's research is supported by the Deutsche Forschungsgemeinschaft (DFG) under Grant No.\ SFB/TRR55.}

\appendix*

\section{The $\xi/\xi_{2nd}$ observable}

   Gauge theories have a complicated spectrum, and correlators at intermediate distances should be sensitive to more than just the lowest-lying excitation. Caselle and Nada \cite{Caselle:2017xrw} have introduced an interesting observable which tests for the presence, in an effective Polyakov line action, of a spectrum of excitations contributing to two-point correlators.  We briefly review their idea in this section, as implemented at $\m=0$.  

   Denoting  $P_\vx=P_{x,y,z}$, define the average of Polyakov lines
in a plane on an $L^3$ lattice volume
\beq
        \P(z) = {1\over L^2} \sum_x \sum_y P_{x,y,z} \ ,
\eeq
and consider the connected correlator\footnote{We adhere in this appendix to the notation of \cite{Caselle:2017xrw}, but $G(\t)$ should not be confused with the usual Polyakov line correlator defined in eq.\ \rf{Pcorr}.}
\beq
         G(\t) = \langle \P(0) \P^\dg(\t) \rangle - |\langle \P \rangle|^2
\label{Gt}
\eeq
We extract the correlation length $\xi$ from a best fit of $G(\t)$ to the form
\beq
          G(\t) \sim c_0 \left( e^{-\t/\xi} + e^{-(L-\t)/\xi} \right)
\label{pfit}
\eeq
which is a form we expect to be true asymptotically for sufficiently large $\t,~ L-\t$.  But suppose that in fact $G(\t)$ is more
accurately expressed as a sum of exponentials, reflecting the existence of excited states, i.e.\ for $\t \ll L/2$
\beq
          G(\t) \approx \sum_i c_i e^{-\t/\xi_i}
\label{Gt1}
\eeq         
Let $\G(\t)$ be $G(\t)$ in the $L \ra \infty$ limit. Following \cite{Caselle:2017xrw} we consider
\beq
          \xi^2_{2nd} = {\sum_{\t=0}^\infty \t^2 \G(\t) \over 2 \sum_{\t=0}^\infty  \G(\t)}
\label{xi2}
\eeq
If we approximate the sums over $\t$ by integrals, and assuming that $\G(\t)$ has the form \rf{Gt} for all $\t$, we find
that
\beq
            \xi^2_{2nd} = {\sum_i c_i \xi_i^3\over \sum_i c_i \xi_i}
\eeq
Now suppose that the sum is dominated by the first term, with all other terms negligible.  In that case $\xi$ defined in \rf{pfit}
equals $\xi_1$, and we have $\xi /\xi_{2nd} = 1$. In the Ising model in $D=3$ dimensions, the deviation from unity is on the order of a few percent. Conversely, if the ratio $\xi/\xi_{2nd}$ is greater than one, then this is evidence that there is a spectrum of excited states, not just the lowest lying excitation, that make a non-negligible contribution to the sum.  For SU(2) pure gauge theory, lattice Monte Carlo simulations well away from the deconfinement transition show a much more substantial deviation, on the order 40-50\% \cite{Caselle:2017xrw}.

 \begin{figure}[h!]
\centerline{\scalebox{0.6}{\includegraphics{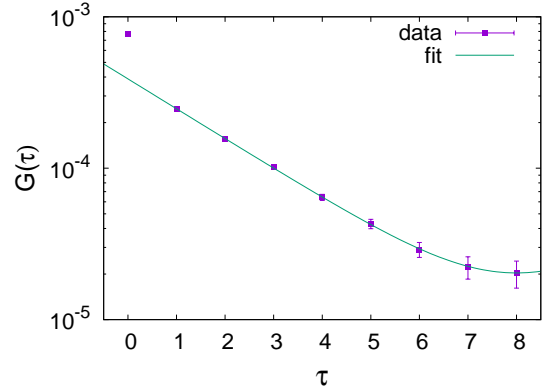}}}
\caption{$G(\t)$ defined in \rf{Gt} vs.\ $\t$ for the PLA corresponding to the LGT at $\b=5.65$ and a $16^3 \times 6$ lattice volume.}
\label{Cas}
\end{figure} 

    On the lattice we have lattice periodicity, and we cannot take the $L\ra \infty$ limit to get $\G(\t)$.  The proposal is instead to approximate \rf{xi2} by
\bea
\lefteqn{\xi^2_{2nd} =} & & \non \\
   & &       { \sum_{t=0}^{\t_{max}} \t^2 G(\t) + \sum_{\t=\t_{max}+1}^\infty \t^2 G(\t_{max}) \exp\left(-{\t-\t_{max} \over \xi}\right) \over
         2 \sum_{t=0}^{\t_{max}}  G(\t) + 2 \sum_{\t=\t_{max}+1}^\infty  G(\t_{max}) \exp\left(-{\t-\t_{max} \over \xi}\right)  } \non \\
\label{approx}
\eea
where $\xi$ is determined from \rf{pfit}.

   There are, of course, possible sources of systematic error in the choice of $\t_{max}$ and the extraction of $\xi$.  But
for a PLA derived for a system of dynamical fermions there is an additional ambiguity.  Polyakov lines at a separation less than
the string-breaking scale represent a quark-antiquark system joined by a flux tube, and the contributions to $G(\t)$ come from a
spectrum of string-like excitations.  Beyond the string-breaking scale, the Polyakov lines represent two bound states, namely a
static quark + light antiquark, and a static antiquark + light quark.  In this case the contributions to the connected correlator
are associated with hadron exchange.  So in this case the ratio $\xi/\xi_{2nd}$ is picking up contributions from quite different 
regimes.  In view of this, we have calculated the ratio $\xi/\xi_{2nd}$ at $\b=5.65$, where the expectation value of the Polyakov
line is very small, and the contributions to $\xi/\xi_{2nd}$ are coming mainly from the string-like spectrum, at least on the comparatively
small lattice volume of $16^4$ used in our simulations. We obtain $\xi= 2.196(11)$ from the fit \rf{pfit}, with data and fit shown in Fig.\ \ref{Cas} shown, and using \rf{approx} with $\t_{max}=5$ we finally obtain
\bea
           {\xi \over \xi_{2nd}} = 1.27 \pm 0.03  \\     \non 
\label{crat}
\eea
which is comparable to some of the results for pure SU(2) lattice gauge theory quoted in ref.\ \cite{Caselle:2017xrw}. 
The fit to a single correlation length \rf{pfit} is expected to be very accurate at large $\t$, and the influence of higher excitations would
only be evident at smaller $\tau$. 
Note that in Fig.\ \ref{Cas} the fit \rf{pfit} is in fact quite accurate for all $\t > 0$, so in this case the deviation of the ratio from unity
must be attributed mainly to the data point at $\t=0$. We must stress again that this result is obtained on a relatively small lattice volume, and is subject to the caveats already mentioned.

\bibliography{pline}

\end{document}